\documentclass[aps, prb, showpacs, twocolumn, amsmath, superscriptaddress]{revtex4}
\usepackage{graphicx}
\usepackage{amssymb}
\begin{document}
\title{Effect of yttrium substitution on the superconducting properties of the system La$_{1-\textit{x}}$Y$_{\textit{x}}$O$_{0.5}$F$_{0.5}$BiS$_{2}$}

\author{I. Jeon}
\affiliation{Department of Physics, University of California, San Diego, La Jolla, California 92093, USA}
\affiliation{Center for Advanced Nanoscience, University of California, San Diego, La Jolla, California 92093, USA}
\affiliation{Materials Science and Engineering Program, University of California, San Diego, La Jolla, California 92093, USA}

\author{D. Yazici}
\affiliation{Department of Physics, University of California, San Diego, La Jolla, California 92093, USA}
\affiliation{Center for Advanced Nanoscience, University of California, San Diego, La Jolla, California 92093, USA}

\author{B. D. White}
\affiliation{Department of Physics, University of California, San Diego, La Jolla, California 92093, USA}
\affiliation{Center for Advanced Nanoscience, University of California, San Diego, La Jolla, California 92093, USA}

\author{A. J. Friedman}
\affiliation{Department of Physics, University of California, San Diego, La Jolla, California 92093, USA}
\affiliation{Center for Advanced Nanoscience, University of California, San Diego, La Jolla, California 92093, USA}

\author{M. B. Maple}
\email[Corresponding Author: ]{mbmaple@ucsd.edu}
\affiliation{Department of Physics, University of California, San Diego, La Jolla, California 92093, USA}
\affiliation{Center for Advanced Nanoscience, University of California, San Diego, La Jolla, California 92093, USA}
\affiliation{Materials Science and Engineering Program, University of California, San Diego, La Jolla, California 92093, USA}

\begin{abstract}
We present the effect of yttrium substitution on superconductivity in the La$_{1-\textit{x}}$Y$_{\textit{x}}$O$_{0.5}$F$_{0.5}$BiS$_{2}$ system. Polycrystalline samples with nominal Y concentrations up to 40\% were synthesized and characterized via electrical resistivity, magnetic susceptibility, and specific heat measurements. Y substitution reduces the lattice parameter \textit{a} and unit cell volume \textit{V}, and a correlation between the lattice parameter \textit{c}, the La-O-La bond angle, and the superconducting critical temperature $T_c$ is observed. The chemical pressure induced by Y substitution for La produces neither the high-$T_c$ superconducting phase nor the structural phase transition seen in LaO$_{0.5}$F$_{0.5}$BiS$_{2}$ under externally applied pressure.
\end{abstract}

\pacs{74.25.F-, 74.62.Dh, 74.62.Fj, 33.15.Dj}

\maketitle

\section{Introduction}
Superconductivity was recently discovered in the layered compound Bi$_4$O$_4$S$_3$ with critical temperature $T_c =$ 8.6 K[\onlinecite{MIZUGUCHI1, AWANA0}]. Shortly thereafter, superconductivity was reported in fluorine-doped \textit{Ln}O$_{1-x}$F$_x$BiS$_2$ (\textit{Ln} = La, Ce, Pr, Nd, Yb) compounds, with a maximum $T_c =$ 10.6 K[\onlinecite{MIZUGUCHI2, DEMURA1, DEGUCHI_DOME, XING1, DUYGU1, AWANA1, AWANA2, AWANA3, AWANA_LA, AWANA_Nd}]. A layered structure is also observed for these materials, composed of superconducting BiS$_2$ and blocking oxide layers. Subsequent studies demonstrated that superconductivity is induced in general by electron doping in the blocking layers, as with the systems \textit{Ln}O$_{1-x}$F$_x$BiS$_2$ (\textit{Ln} = La, Ce, Pr, Nd, Yb) and La$_{1-x}$$M_x$OBiS$_2$ (\textit{M} = Ti, Zr, Hf, Th)[\onlinecite{DUYGU2}], or doping in the isocharge block [$Ln_2$O$_2$]$^{2-}$ with a [Sr$_2$F$_2$]$^{2-}$ layer[\onlinecite{XI1,YUKE1}]. The parent compounds, \textit{A}OBiS$_2$ (\textit{A} = La, Ce, Th)[\onlinecite{DUYGU2,XING1}] and SrFBiS$_2$[\onlinecite{XI1,YUKE1, SAKAI1}], are bad metals and show semiconducting-like behavior; however, theoretical studies employing the tight-binding model and density functional calculations predicted that electron doping in the BiS$_2$ system increases the density of states at the Fermi level[\onlinecite{USUI1,WAN1}], making electron doping a crucial tuning parameter for superconductivity. The pairing mechanism in both Bi$_4$O$_4$S$_3$ and LaO$_{0.5}$F$_{0.5}$BiS$_2$ has also been investigated; recent studies of the temperature dependence of the penetration depth by the tunnel diode oscillator technique revealed evidence for fully gapped, strongly-coupled s-wave superconductivity in the Bi$_4$O$_4$S$_3$ compound[\onlinecite{SHRUTI1}], and an s-wave character for LaO$_{0.5}$F$_{0.5}$BiS$_2$ was indicated in muon-spin spectroscopy measurements[\onlinecite{LAMURA1}]. It has been suggested that superconductivity emerges in the vicinity of a charge-density wave (CDW) and semiconducting-like behavior[\onlinecite{XING1,YILDIRIM1}]. Moreover, neutron scattering measurements on the LaO$_{1-x}$F$_{x}$BiS$_2$ system show intrinsic structural instabilities in the superconducting phases[\onlinecite{LEE1}]. As a consequence, studies focused on applied pressure as a tuning parameter in BiS$_2$ compounds have been conducted recently. It has been reported that the \textit{Ln}O$_{0.5}$F$_{0.5}$BiS$_2$ (\textit{Ln} = La, Ce, Pr, Nd) compounds show marked $T_c$ enhancements[\onlinecite{KOTEGAWA1, SELVAN1, SELVAN2, WOLOWIEC1, WOLOWIEC2}] when subjected to applied pressure. A study of LaO$_{0.5}$F$_{0.5}$BiS$_2$ with various lattice parameters has shown that reducing the lattice parameters should have an effect on $T_c$[\onlinecite{KAJITANI1}]. To further investigate the relationships between pressure, lattice parameters, and superconductivity, chemical substitution of Y for La is a logical way to tune the properties of LaO$_{0.5}$F$_{0.5}$BiS$_2$. Like La, Y has no magnetic moment and a trivalent electronic configuration, and chemical pressure can be introduced by partial substitution of La by smaller Y ions. The effect of Y substitution has been studied for a number of superconducting systems, with suppression of superconductivity observed in the systems (La$_{1-x}$Y$_x$)$_{1.85}$Sr$_{0.15}$CuO$_4$ and (La$_{1-x}$Y$_x$)NiC$_2$[\onlinecite{LAY1,LAY2}], and enhancement of superconductivity observed in (La$_{1-x}$Y$_x$)Co$_2$B$_2$, La$_{1-x}$Y$_x$FeAsO$_{1-\delta}$, and F-doped La$_{1-y}$Y$_y$FeAsO[\onlinecite{LAY3,SHIRAGE1,TROPEANO1}]. The latter system shows a remarkable enhancement of $T_c$ from 24 K to 40 K. The effect of Y substitution on the BiS$_2$ systems has not been explored yet. In this work, we present a systematic study in which we have substituted Y ions into the La-site in LaO$_{0.5}$F$_{0.5}$BiS$_{2}$. We observe that the critical temperature $T_c$ appears to be correlated with the lattice parameter $c$ and the La-O-La bond angle. The chemical pressure resulting from Y substitution is insufficient to induce the structural phase transition from tetragonal (\textit{P}4/\textit{nmm}) to monoclinic (\textit{P}2$_1$/\textit{m}) crystal structures seen under an applied pressure of 1 GPa in LaO$_{0.5}$F$_{0.5}$BiS$_{2}$[\onlinecite{TOMITA1}].

\section{Experimental Details}
Polycrystalline samples of La$_{1-x}$Y$_{x}$O$_{0.5}$F$_{0.5}$BiS$_{2}$ (0 $\leq$ \textit{x} $\leq$ 0.40) were prepared by a conventional solid state reaction method. High-purity starting materials (purity $\geq$ 99.9\%) of La, Y, and S, as well as LaF$_{3}$, Bi$_{2}$O$_{3}$, and Bi$_{2}$S$_{3}$ were weighed stoichiometrically. They were well-mixed, pressed into pellets, encapsulated in evacuated quartz tubes, and annealed at 800 $^\circ$C for two days. This process was repeated two additional times to promote homogeneity of the samples. The crystal structure was determined by x-ray powder diffraction (XRD) using a Bruker D8 Discover x-ray diffractometer with Cu-K$_{\alpha}$ radiation and XRD patterns were analyzed via Rietveld refinement using the GSAS+EXPGUI software package[\onlinecite{GSAS1,GSAS2}]. The temperature dependence of electrical resistivity was measured from 1.1 K to 300 K using a standard four-wire method with a Linear Research LR700 ac resistance bridge and a home-built probe in a liquid $^4$He Dewar. Magnetic susceptibility measurements were performed between 2 K and 10 K with applied magnetic field $H$ = 5 Oe using a Quantum Design Magnetic Properties Measurement System (MPMS). Alternating current magnetic susceptibility was measured down to $\sim$ 1.1 K in a liquid $^4$He Dewar using home-built magnetic susceptibility coils. Specific heat measurements were made for 1.8 K $\leq$ \textit{T} $\leq$ 30 K with a Quantum Design Physical Properties Measurement System (PPMS) DynaCool.

\begin{figure}
\includegraphics[width=8.5cm]{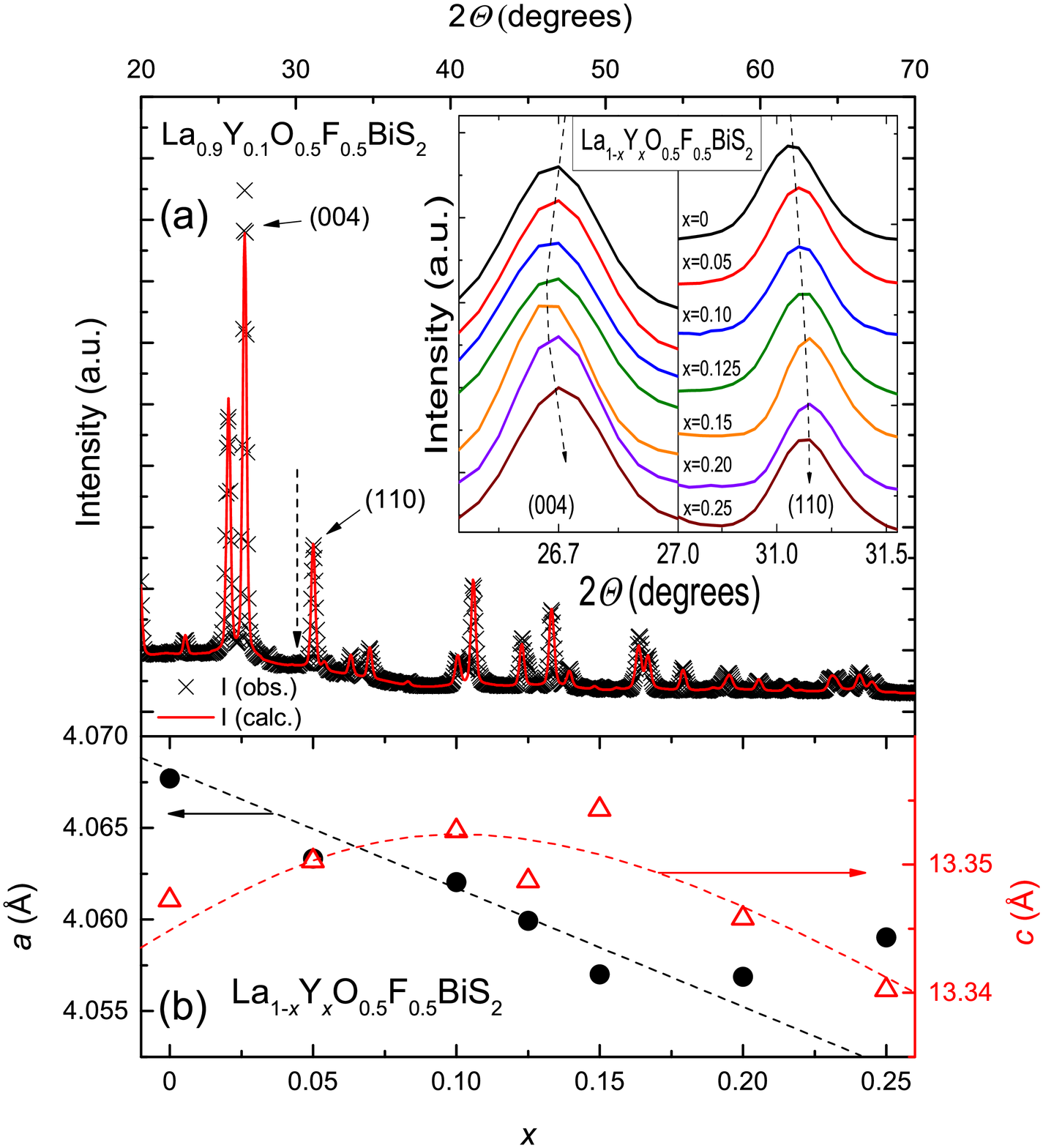}
\caption{
(Color online) (a) X-ray diffraction pattern for La$_{0.9}$Y$_{0.1}$O$_{0.5}$F$_{0.5}$BiS$_{2}$. The black crosses are data and the red line represents the fit results from Rietveld refinement of the data. The dashed arrow indicates a La$_{2}$O$_{2}$S and/or Y$_{2}$O$_{2}$S impurity. The systematic behavior of the (004) and (110) diffraction peaks is shown in the inset of the graph. (b) Lattice parameters $a$ and $c$ versus nominal yttrium concentration \textit{x}.
}
\label{fig:XRD}
\end{figure}

\section{Results}

\subsection{X-ray diffraction}
\begin{figure}
\includegraphics[width=8.5cm]{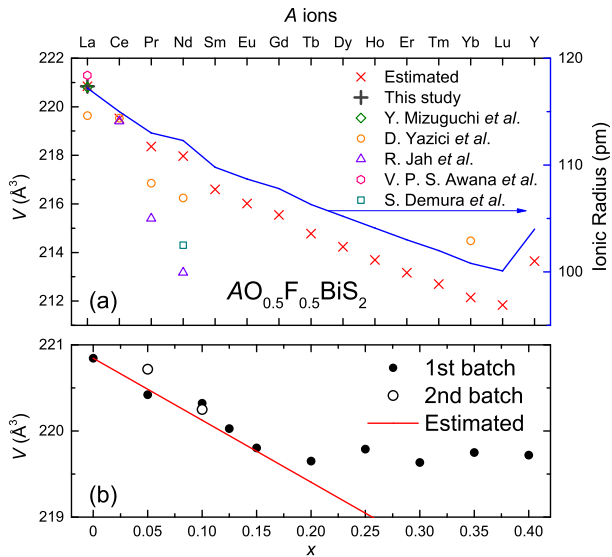}
\caption{
(Color online) (a) Estimated volumes and ionic radii versus lanthanide and Y ions are shown for $A$O$_{0.5}$F$_{0.5}$BiS$_2$. The blue line is the trend of ionic radii for lanthanide and Y ions. Red crosses are estimated volumes and other symbols are reported volumes for $Ln$O$_{0.5}$F$_{0.5}$BiS$_2$ ($Ln =$ La, Ce, Pr, Nd, Yb). The estimated volume of YO$_{0.5}$F$_{0.5}$BiS$_2$ is at the right corner. (b) Black circles and red line are observed and estimated unit cell volumes \textit{V} versus $x$, respectively. Unfilled circles correspond to a second batch of samples prepared for selected Y concentrations.
}
\label{fig:XRD2}
\end{figure}

Figure~\ref{fig:XRD} shows XRD data for La$_{1-\textit{x}}$Y$_{\textit{x}}$O$_{0.5}$F$_{0.5}$BiS$_{2}$ (0 $\leq$ \textit{x} $\leq$ 0.25) samples. All XRD patterns are well indexed by the tetragonal CeOBiS$_{2}$-type crystal structure with space group \textit{P}4/\textit{nmm}. Figure~\ref{fig:XRD} (a) displays the XRD pattern and the result of Rietveld refinement of the data for the La$_{0.9}$Y$_{0.1}$O$_{0.5}$F$_{0.5}$BiS$_{2}$ sample. The dashed arrow indicates the presence of La$_{2}$O$_{2}$S and/or Y$_{2}$O$_{2}$S impurity phases, the amount of which increases gradually with increasing $x$. This implies a possible minor discrepancy between nominal and actual yttrium concentrations. The samples with $x \leq$ 0.25 contain the same impurity phase constituting 1$\%$-8$\%$ of the sample by mass and less than 1$\%$ of possible Y and Bi/Bi$_2$S$_3$ impurity phases by mass, as estimated by Rietveld refinements. The systematic behavior of the (004) and (110) diffraction peaks is shown in the inset of Fig.~\ref{fig:XRD} (a) and the $a$ and $c$ lattice parameters are plotted as a function of nominal Y concentration in Fig.~\ref{fig:XRD} (b).\\
\indent To estimate the true Y concentration in our samples, we calculated the expected unit cell volume of YO$_{0.5}$F$_{0.5}$BiS$_2$ (which has thus far not been successfully synthesized) and then compared the measured volumes of our La$_{1-\textit{x}}$Y$_{\textit{x}}$O$_{0.5}$F$_{0.5}$BiS$_{2}$ samples against the expected behavior from Vegard's law. To estimate the unit cell volume of YO$_{0.5}$F$_{0.5}$BiS$_2$, we first calculated the total volume of the ions residing in a single unit cell of LaO$_{0.5}$F$_{0.5}$BiS$_2$ using ionic radii values of the elements from Ref. [\onlinecite{SHANNON}]. We then computed a scale factor by comparing the total volume of the ions with the measured unit cell volume of LaO$_{0.5}$F$_{0.5}$BiS$_2$. Assuming a similar scale factor is appropriate for the compounds containing other rare-earth ions, we made similar calculations to estimate their unit cell volumes. The trend of these estimated unit cell volumes for rare-earth ions is displayed in Fig.~\ref{fig:XRD2} (a) along with measured values for some compounds. Our estimate seems to work particularly well for CeO$_{0.5}$F$_{0.5}$BiS$_2$, but we note that there is a significant spread in experimentally-measured unit cell volumes for the other compounds and that our estimates are reasonable given such uncertainty. Invoking Vegard's law, we plot a line between the measured unit cell volume for LaO$_{0.5}$F$_{0.5}$BiS$_2$ and our estimated value for the volume of YO$_{0.5}$F$_{0.5}$BiS$_2$, and compare this line with the measured volumes of our La$_{1-\textit{x}}$Y$_{\textit{x}}$O$_{0.5}$F$_{0.5}$BiS$_{2}$ samples up to $x =$ 0.40 in Fig.~\ref{fig:XRD2} (b). The agreement between measured and estimated volumes is good up to $x =$ 0.20 and this simple procedure helps to get an idea of the uncertainty for the Y concentrations; the difference between nominal and estimated concentrations can be as large as 3\% for the first batch of the $x =$ 0.10 sample, for instance.\\
\indent For $x \geq$ 0.25, the unit cell volumes $V$ are concentration-independent, as seen in Fig.~\ref{fig:XRD2} (b) and the XRD patterns contain impurity phases of Y, La$_{2}$O$_{2}$S/Y$_{2}$O$_{2}$S, and Bi/Bi$_{2}$S$_{3}$, suggesting that the sample with $x =$ 0.25 is near or even beyond the solubility limit. Therefore, we can conservatively conclude that Y is incorporated into the La site up to $x =$ 0.20 in this system. Since the system forms with a tetragonal crystal structure (\textit{P}4/\textit{nmm}), the lattice parameter $a$ has a more dominant effect than the lattice parameter $c$ on the unit cell volume $V$, with both $a$ and $V$ decreasing with increasing $x$. Also, the ionic radius of Y is less than that of La, suggesting that chemical pressure is induced in La$_{1-x}$Y$_x$O$_{0.5}$F$_{0.5}$BiS$_{2}$ up to $x =$ 0.20.

\begin{figure}
\includegraphics[width=9cm]{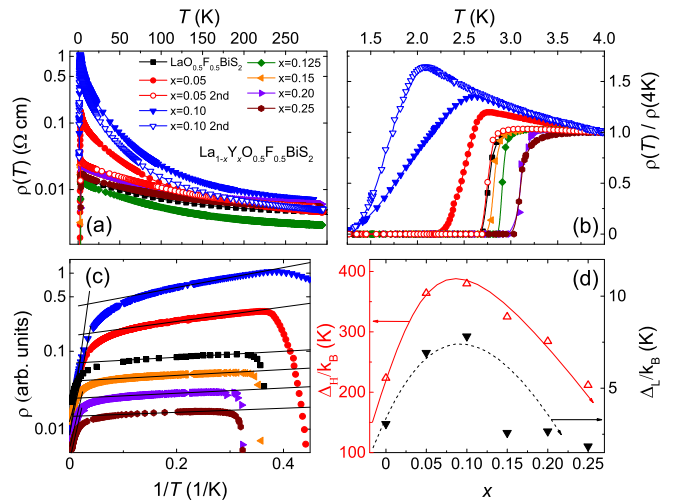}
\caption{
(Color online) (a) A semilogarithmic plot of electrical resistivity $\rho$ versus temperature \textit{T}. (b) $\rho$, normalized to its value at 4 K, versus temperature \textit{T}. (c) ln $\rho$ versus 1/\textit{T} for selected samples with \textit{x} = 0, 0.05, 0.10, 0.15, 0.20, and 0.25. The solid lines indicate simple activation-type temperature dependencies from Eq.~(\ref{eq:Egap}). The data are vertically offset for visual clarity. (d) Energy gaps $\Delta_{H}$/$k_B$ and $\Delta_{L}$/$k_B$ calculated at high and low temperatures respectively, versus nominal yttrium concentration \textit{x}.
}
\label{fig:RHO}
\end{figure}

\subsection{Electrical Resistivity}
Electrical resistivity $\rho$(\textit{T}) data are plotted in Fig.~\ref{fig:RHO}. For all samples in their normal states, electrical resistivity exhibits semiconducting-like behavior and clear drops at the superconducting transition temperature \textit{T}$_{c}$. We determined $T_c$ by measuring the temperatures where the electrical resistivity falls to 50\% of its normal-state value, and the broadness of the transitions was characterized by identifying the temperatures where the electrical resistivity decreases to 90\% and 10\% of the normal-state value. We observe two different types of behavior: for $x \leq$ 0.10, $\rho$ increases rapidly with decreasing temperature in its normal-state and a broad superconducting transition is observed, while a slower increase of $\rho$ with decreasing temperature accompanying a sharp superconducting transition is seen for $x >$ 0.10. Such behavior is emphasized by plotting $\rho$(\textit{T}), normalized by its value at 4 K, versus temperature \textit{T} in Fig.~\ref{fig:RHO} (b). To confirm the reproducibility of these different types of behavior, we have synthesized and characterized several additional samples for each Y concentration, especially for $x =$ 0.05 and 0.10. Since we observed the same behavior for different samples within all batches, shown in Figs.~\ref{fig:XRD2}(b), ~\ref{fig:RHO}(a) and (b), and ~\ref{fig:Tc}(b), we only present representative data for each concentration in this study. To estimate the energy gaps, we adopted the simple activation-type relation[\onlinecite{KOTEGAWA1}],
\begin{eqnarray}\label{eq:Egap}
\rho(T) = \rho_0e^{\Delta/2k_{B}T} ,
\end{eqnarray}
where $\rho_0$ is a constant and $\Delta$ is the energy gap. As shown in Fig.~\ref{fig:RHO} (c), we fit Eq.~\ref{eq:Egap} to data for selected samples in two regions, 200 - 300 K and \textit{T}$_{c}$ - 20 K, to obtain a high-temperature energy gap $\Delta_{H}$/$k_B$ and a low-temperature energy gap $\Delta_{L}$/$k_B$, respectively
. Both energy gaps are found to first increase with $x$ up to $x =$ 0.10 and then decrease with higher concentration, as illustrated in Fig.~\ref{fig:RHO} (d). This behavior exhibits the same trend as that of the lattice parameter $c$. The behavior of the electrical resistivity up to $x =$ 0.10 is different from that observed under applied external pressure on BiS$_{2}$-based superconducting compounds[\onlinecite{KOTEGAWA1, SELVAN1, SELVAN2, WOLOWIEC1, WOLOWIEC2, TOMITA1}], in which semiconducting-like behavior is suppressed with increasing pressure and a metallic state is induced. On the other hand, we note that there is currently no data for applied pressures less than 0.3 GPa, which is larger than the chemical pressure in the $x =$ 0.10 compound (as will be discussed later). The semiconducting-like behavior is gradually suppressed for $x \geq$ 0.10, which is similar to the results reported in pressure studies[\onlinecite{WOLOWIEC1, WOLOWIEC2}].

\begin{figure}
\includegraphics[width=8.5cm]{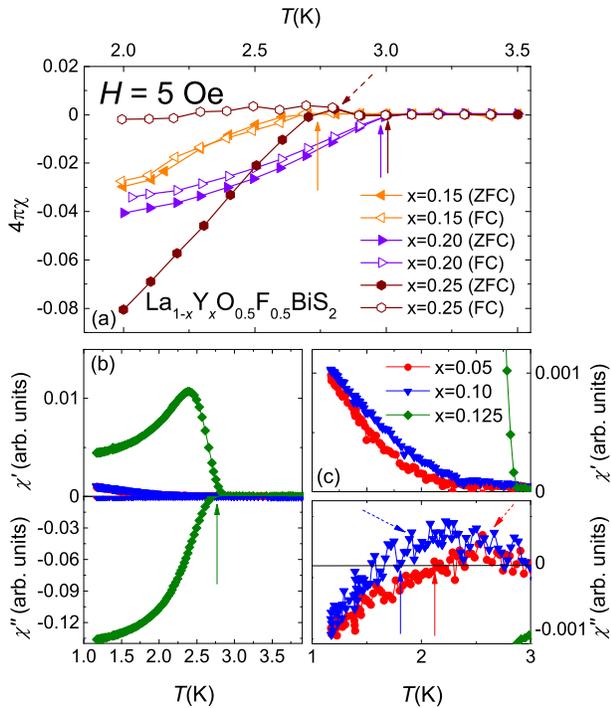}
\caption{
(Color online) (a) Magnetic susceptibility $\chi_{dc}$ versus temperature \textit{T} for La$_{1-\textit{x}}$Y$_{\textit{x}}$O$_{0.5}$F$_{0.5}$BiS$_{2}$ (\textit{x} = 0.15, 0.20, and 0.25), measured in field cooled and zero-field cooled conditions. The solid arrows denote the superconducting critical temperature $T_c$ and the dashed arrow emphasizes the presence of a small upturn. (b) and (c) Alternating current magnetic susceptibility $\chi_{ac}$ versus temperature $T$ for $x=$ 0.05, 0.10, and 0.125. The solid and dashed arrows have the same meaning as in panel (a). 
}
\label{fig:CHI}
\end{figure}

\begin{table}[t]
\caption{Impurity phases La$_2$O$_2$S and/or Y$_2$O$_2$S by mass and superconducting volume fraction at 2 K along with $T_c$ from $\rho$ data for each concentration.}        
\centering   
    \begin{tabular}{ l | c | c | c | c }
    \hline
    \hline
    ~~~~~\textit{x} ~~~~~& ~~$T_c$ (K)~~& ~Imp. (\%) & ~$V$ frac. (\%) &Ref.\\ 
    \hline
    ~Parent	& 2.7  & -& $\sim$ 6&[\onlinecite{AWANA_LA}]\\
    & 3.0& -& ~$\sim$ 13&[\onlinecite{MIZUGUCHI2}]\\
    & 3.1 & -& ~$\sim$ 60&[\onlinecite{DUYGU1}]\\
    & 2.8 & $< $ 1 &- &[This study]\\
    \hline
    ~~~0.05	& 2.5  & $\sim$ 1 &~$\ll$ 1  &\\
    ~~~0.10	& 1.8  & $\sim$ 2 &~$\ll$ 1 &\\
    ~~~0.125	& 2.9  & $\sim$ 2 &~~$\sim$ 11 & [This study]\\
    ~~~0.15	& 2.8  & $\sim$ 3 &~$\sim$ 3  &\\
    ~~~0.20	& 3.1  & $\sim$ 4 &~$\sim$ 4  &\\
    ~~~0.25	& 3.1  & $\sim$ 8 &~$\sim$ 8  &\\
    \hline
    \hline
    \end{tabular}
    \label{tbl:VF}
    \end{table}

\subsection{Magnetic Susceptibility}
In order to characterize the observed superconductivity in the La$_{1-\textit{x}}$Y$_{\textit{x}}$O$_{0.5}$F$_{0.5}$BiS$_{2}$ system, temperature-dependent dc magnetic susceptibility measurements were performed under a 5 Oe magnetic field with both zero-field-cooled (ZFC) and field-cooled (FC) methods, and the results are displayed in Fig.~\ref{fig:CHI} (a). We performed measurements on selected samples with \textit{x} = 0.15, 0.20, and 0.25 which have superconducting transition temperatures $T_c$ in the accessible temperature range of the MPMS. Clear diamagnetic signals were observed for each of these samples. The \textit{T}$_{c}$ values, determined by the temperatures at the onset of the diamagnetic signal, are indicated by the solid arrows in Fig.~\ref{fig:CHI} (a) and are in good agreement with those estimated from the electrical resistivity data. Alternating current magnetic susceptibility measurements for the samples with $x =$ 0.05, 0.10, and 0.125 are shown in Figs.~\ref{fig:CHI} (b) and (c). A clear signature of superconductivity was observed for $x =$ 0.125 and the onset of a SC signal was observed for the $x =$ 0.05, 0.10 samples. Though the transitions are not complete, we estimated superconducting shielding fractions of $\sim$ 13\% at 1.1 K for the $x =$ 0.125 sample and $\sim$ 3\%, $\sim$ 4\%, and $\sim$ 8\% at 2 K for the \textit{x} = 0.15, 0.20, and 0.25 samples, respectively. The weak signals for $x =$ 0.05, 0.10 followed by increasing shielding fractions for $x \geq$ 0.125 seem to correlate with the width of the transitions and the variation of the $T_c$ values observed in electrical resistivity measurements. These shielding fractions are similar to reported values of $\sim$ 13\% and $\sim$ 6\%[\onlinecite{MIZUGUCHI2,AWANA_LA}], but smaller than that of the highest value of $\sim$ 60\%[\onlinecite{DUYGU1}]. The transition temperatures $T_c$ from $\rho$, the amount of La$_2$O$_2$S and/or Y$_2$O$_2$S impurity phases by mass, and superconducting volume fractions at 2 K are summarized in Table~\ref{tbl:VF}. These results imply that the volume fractions do not seem to correlate with Y concentrations or the amount of impurity phases; volume fractions more likely correlate with the transition temperatures and the broadness of transitions. The samples with $x =$ 0.05, 0.10, and 0.25 exhibit a weak upturn indicated by dashed arrows in dc and ac susceptibility data in Fig.~\ref{fig:CHI}. This behavior is probably due to a small amount of paramagnetic impurities, as observed in other studies[\onlinecite{DUYGU1,DUYGU2,AWANA1,AWANA2,AWANA3}].

\begin{figure}
\includegraphics[width=8.5cm]{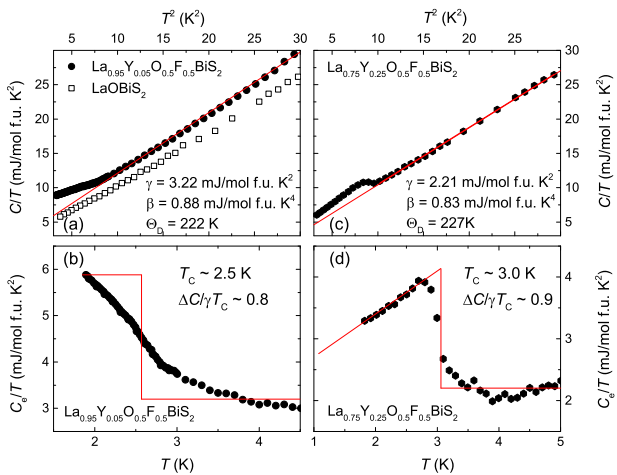}
\caption{
(Color online) (a) and (c) Specific heat divided by temperature $C/T$ versus temperature \textit{T} for LaOBiS$_{2}$, La$_{0.95}$Y$_{0.05}$O$_{0.5}$F$_{0.5}$BiS$_{2}$, and La$_{0.75}$Y$_{0.25}$O$_{0.5}$F$_{0.5}$BiS$_{2}$. The red lines represent the best fits of the equation $C(T)/T=\gamma + \beta T^{2}$ to the data which yield $\gamma =$ 3.22 mJ/mol f.u.K$^{2}$ and $\Theta _{D}$ = 222 K for $x =$ 0.05 and $\gamma =$ 2.21 mJ/mol f.u.K$^{2}$ and $\Theta _{D}$ = 227 K for $x =$ 0.25. (b) and (d) Plots of the electronic contribution to the specific heat divided by $T$, $C_e/T$, versus temperature \textit{T}. Idealized entropy conserving constructions result in an estimate of $\Delta C/\gamma T_c =$ 0.8 for $x =$ 0.05 if we use $T_c = $ 2.5 K from electrical resistivity measurements (see text) and $T_c = $ 3.0 K and $\Delta C/\gamma T_c =$ 0.9 for $x =$ 0.25.}
\label{fig:Cp}
\end{figure}

\subsection{Specific Heat}
To further investigate the superconducting properties of this system, measurements of the specific heat, $C$, were performed for the $x=$ 0.05 and 0.25 samples in the temperature range from 1.8 K to 30 K. The results of these measurements are displayed in Fig.~\ref{fig:Cp}. Although the specific heat jump for the $x = $ 0.05 sample is incomplete, possibly due to its low $T_c$, we observed a broad upturn which is consistent with the $T_c$ values obtained from $\rho$ (2.5 K) and $\chi_{ac}$ (2.2 K), suggesting that this feature is assocaited with superconductivity. For the $x =$ 0.25 sample, a clear specific heat jump is observed at $T_c \simeq$ 3.0 K, in good agreement with the $T_c$ values from $\rho$ (3.1 K) and $\chi_{dc}$ (3.0 K). The appearance of a jump in $C/T$ at $T_c$ is strong evidence that the superconductivity for $x =$ 0.05 and 0.25 is a bulk phenomenon. The specific heat at low temperature can be written,
\begin{eqnarray}
C(T) = \gamma T + \beta T^3 ,
\label{eq:Cp}
\end{eqnarray}
where the terms $\gamma T$ and $\beta T^3$ account for the electronic and phonon contributions, respectively. The data were fitted to this expression, yielding the electronic coefficient $\gamma =$ 3.22 mJ/mol f.u.K$^2$ and the lattice coefficient $\beta =$ 0.88 mJ/mol f.u.K$^4$ for the $x =$ 0.05 sample and $\gamma =$ 2.21 mJ/mol f.u.K$^2$ and $\beta =$ 0.83 mJ/mol f.u.K$^4$ for the $x =$ 0.25 sample. The best fits, shown in Figs.~\ref{fig:Cp} (a) and (c), yield the fitting parameters listed in the figures. To obtain the ratios of the specific heat jump to $\gamma T_c$, the lattice contributions were subtracted revealing the upturn and specific heat jump in Fig.~\ref{fig:Cp} (b) and (d), respectively. Since the jump was incomplete for the $x =$ 0.05 sample, a rough estimate of the ratio $\Delta C/\gamma T \sim$ 0.8 was extracted from the size of the upturn and assuming $T_c \sim$ 2.5(1) K, as illustrated in Fig.~\ref{fig:Cp} (b). This value is comparable to the value of 0.9 for the $x =$ 0.25 sample, and both of these are smaller than the value of 1.43 predicted by the weak-coupling Bardeen-Cooper-Schrieffer (BCS) theory of superconductivity; on the other hand, our values of $\gamma$, $\Theta _D$, and $\Delta C/\gamma T$ are similar to $\gamma =$ 2.53 mJ/mol f.u.K$^2$, $\Theta _D =$ 221 K, and $\Delta C/\gamma T =$ 0.94 reported for LaO$_{0.5}$F$_{0.5}$BiS$_{2}$[\onlinecite{DUYGU1}].\\
\indent It is noteworthy that the electronic specific heat behaves so differently from the predictions of the BCS theory. According to the BCS weak-coupling limit, the electronic specific heat below $T_c$ decreases exponentially with decreasing temperature and almost reaches zero near $T_c /$5[\onlinecite{MGB2}]. In contrast to the BCS limit, our data for $x =$ 0.25 are still larger than the normal state value at $T_c /$2, as seen in Fig.~\ref{fig:Cp} (d). Similar behavior of specific heat was observed in previous studies[\onlinecite{DUYGU2, XI1}]. Attempts to fit the specific heat data for $x =$ 0.25 to a simple BCS expression for the low-temperature electronic specific heat, $C_e / \gamma T_c = 1.34(\Delta (0) / T)^{3/2} e^{-\Delta (0) / T}$[\onlinecite{MERMIN}], where the exponential drop is determined by the zero-temperature energy gap, $\Delta (0)$, required the inclusion of additional temperature dependent and constant terms to satisfactorily fit the data (data and fits not shown). It is already known that these samples are not completely homogeneous, and such behavior suggests that part of the sample behaves as a bulk superconductor while the rest (the impurity phase portions of the sample) provide a non-superconducting contribution to $C/T$. The combination of these contributions presumably leads to the distinctly non-BCS temperature dependence we have observed in the behavior of $C_e (T)/T$ below $T_c$. We also note that there is a debate regarding whether or not BiS$_2$-based superconductors actually exhibit conventional BCS superconductivity: weak electron-phonon coupling, a rather large value of 2$\Delta / k_B T_c \sim$ 17, giant superconducting fluctuations, and an anomalous semiconducting normal state have been considered for these compounds[\onlinecite{LEE1, NOBCS, HHWEN1}]. More investigations will be needed to determine the mechanism and nature of superconductivity in BiS$_2$-based compounds.

\begin{figure}
\includegraphics[width=8.5cm]{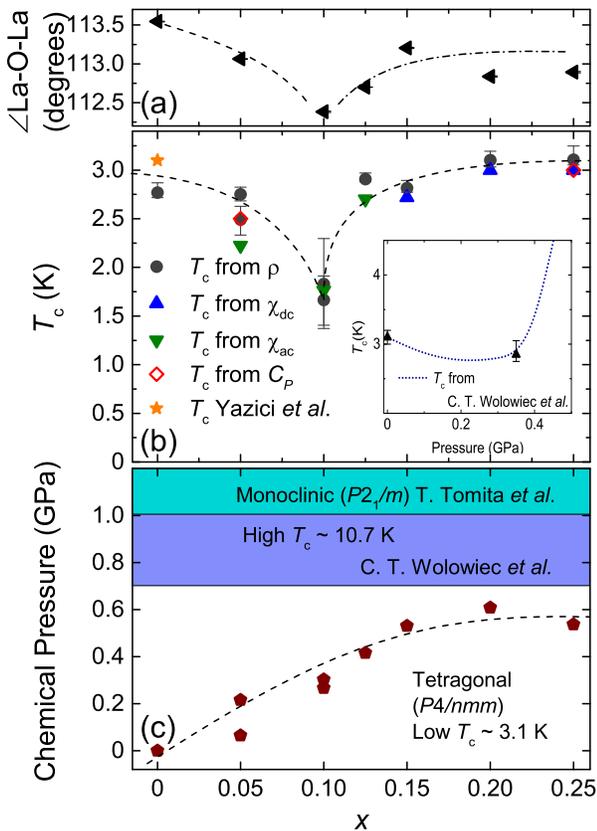}
\caption{
(Color online) (a) La/Y-O/F-La/Y bond angle $\angle$La-O-La versus nominal yttrium concentration $x$. (b) Superconducting transition temperature $T_c$ versus nominal yttrium concentration \textit{x} for the La$_{1-x}$Y$_{x}$O$_{0.5}$F$_{0.5}$BiS$_{2}$ system. The orange star is the maximum $T_c$ obtained from electrical resistivity measurements of LaO$_{0.5}$F$_{0.5}$BiS$_{2}$ reported in Ref. [\onlinecite{DUYGU1}]. The inset shows the behavior of $T_c$ under applied pressure[\onlinecite{WOLOWIEC2}]. (c) Chemical pressure versus nominal yttrium concentration. Two regions with low $T_c$ and high $T_c$ (purple and blue regions) are separated by the reported critical pressure value $P_c\sim$ 0.7 GPa[\onlinecite{KOTEGAWA1,WOLOWIEC1,TOMITA1}]. The tetragonal (unfilled and purple regions) and monoclinic (blue region) phases are distinguished by the critical pressure $P_T\sim$ 1 GPa for the structural phase transition[\onlinecite{TOMITA1}]. 
}
\label{fig:Tc}
\end{figure}

\section{Discussion}
We summarize the results from $\rho$, $\chi_{dc}$, $\chi_{ac}$, and $C$ measurements in a phase diagram of transition temperature $T_c$ versus nominal yttrium concentration \textit{x}, as shown in Fig.~\ref{fig:Tc} (b). $T_c$ decreases from 2.8 K to 1.8 K with increasing \textit{x} until $x=$ 0.10 and is roughly constant with a value of $\sim$ 3.0 K for $x\geq$ 0.125.
$T_c$ was found to decrease slightly in a study of the LaO$_{0.5}$F$_{0.5}$BiS$_2$ system under low applied pressures[\onlinecite{WOLOWIEC1,WOLOWIEC2}], as shown in the inset of Fig.~\ref{fig:Tc} (b). This behavior of $T_c(P)$ resembles that of $T_c(x)$ in La$_{1-x}$Y$_x$O$_{0.5}$F$_{0.5}$BiS$_2$. This observation probably indicates that there is a relationship between $T_c$ and crystal structure details. With that possibility in mind, we found that $T_c(x)$ is related to the lattice constant $c$ (see Fig.~\ref{fig:XRD} (b)) and the La-O-La bond angle (see Fig.~\ref{fig:Tc} (a)).\\
\indent Since the unit cell volume \textit{V} decreases with increasing Y concentration until $x =$ 0.20, the variation of $T_c$ could be discussed in the context of chemical pressure. We adopted a value of the isothermal compressibility[\onlinecite{TOMITA1}], $-d(V/V_0)/dP=$ 0.0089 GPa$^{-1}$ (bulk modulus is 112 GPa), where $V$ and $V_0$ are the unit cell volumes with and without Y, respectively. A graph of chemical pressure versus the nominal Y concentration \textit{x} is plotted in Fig.~\ref{fig:Tc} (c) and the high $T_c$ transition pressure $P_c\sim$ 0.7 GPa[\onlinecite{KOTEGAWA1, WOLOWIEC1}] and structural phase transition pressure $P_T\sim$ 1 GPa[\onlinecite{TOMITA1}] are illustrated as purple and blue regions, respectively. Since LaO$_{0.5}$F$_{0.5}$BiS$_2$ with the tetragonal crystal structure (\textit{P}4/\textit{nmm}) is stable up to $\sim$ 0.8 GPa and then experiences a complete structural phase transition above $\sim$ 1.5 GPa, as discussed in previous studies of Tomita \textit{et al.} and Mizuguchi \textit{et al.}[\onlinecite{TOMITA1, MIZUGUCHI2}], the monoclinic phase is presumably responsible for the high-$T_c$ superconducting phase. Our results show that chemical pressure increases with Y concentrations 0 $\leq x \leq$ 0.20 and saturates at a value of $\sim$ 0.6 GPa at the solubility limit. In Fig.~\ref{fig:Tc} (c), it is clear that the induced chemical pressure is insufficient to induce the high-$T_c$ or the monoclinic phase. If the chemical pressure could be further increased, we expect that the high-$T_c$ and monoclinic phases would be induced for $x \geq 0.20$. This is a simple explanation for why this system did not exhibit an enhancement of superconductivity.\\
\indent Chemical pressure alone is unable to account for the suppression of superconductivity for $x \leq$ 0.10. The lattice parameter $c$ shows different behavior from the unit cell volume $V$. It increases slowly with increasing $x$ to $x =$ 0.10 and then decreases for $x \geq$ 0.15 (see Fig.~\ref{fig:XRD} (b)), in contrast to the monotonic increase of chemical pressure to $x =$ 0.20. One possible scenario is that the La/Y-O/F-La/Y ($\angle$La-O-La) bond angle evolves with $x$. As shown in Fig.~\ref{fig:Tc} (a), we observed a decrease of the bond angle for 0 $\leq x \leq$ 0.10 and then an increase for 0.125 $\leq x \leq$ 0.25, as obtained in our Rietveld refinements. Under applied pressure, the lattice parameters $a$ and $c$ of LaO$_{0.5}$F$_{0.5}$BiS$_2$ decrease continuously until a structural phase transition is induced near 1 GPa[\onlinecite{TOMITA1}]. However, chemical pressure is insufficient to induce the structural phase transition. The suppression and subsequent enhancement of superconductivity with $x$ are probably related to the variation of the lattice parameter \textit{c}, suggesting that the superconducting transition temperature is tuned by $c$ in BiS$_2$-based systems, which is consistent with the conclusions of another recent experimental study[\onlinecite{KAJITANI2}]. The width of the superconducting transition, energy gap values, and superconducting volume fractions also seem to vary systematically with the lattice parameter $c$.\\
\indent To the best of our knowledge, studies on BiS$_2$-based compounds still report difficulties synthesizing homogeneous samples; superconducting critical temperatures for the same systems show perceptible discrepancies between studies[\onlinecite{MIZUGUCHI2,XING1,DEMURA1,AWANA1,AWANA2,AWANA3,DUYGU1}]. Also, the lattice parameters or volumes of the same compounds vary for different studies, as seen in Fig.~\ref{fig:XRD2}(a). Often, the systematic chemical substitution studies of the same systems have different phase diagrams[\onlinecite{XING1, AWANA1,DEMURA1,AWANA_Nd, YUKE1, SAKAI1}]. Sometimes $T_c$ does not change and seems to be independent of substituent concentrations until a solubility limit emerges, even though both parent compounds are stable and can be synthesized[\onlinecite{DUYGU2}]. \\
\indent These discrepancies in the lattice parameters, phase diagrams, and transition temperatures $T_c$ between studies might have one or more possible causes: Fluorine substitution studies for \textit{Ln}O$_{1-x}$F$_x$BiS$_2$ (\textit{Ln} = La, Ce, Nd)[\onlinecite{MIZUGUCHI2,XING1, DEMURA1}] showed variations of lattice parameters and $T_c$ for similar nominal fluorine concentrations. Because of the quantitative inaccuracy of EDX measurements for these materials[\onlinecite{HHWEN1}], it is difficult to estimate the exact amount of fluorine in BiS$_2$ compounds. Thus, it is probable that the actual and nominal fluorine ratio could be different, resulting in differences of lattice parameters and $T_c$. On the other hand, recent studies of angle-resolved photoemission spectroscopy (ARPES) and optical spectroscopy[\onlinecite{ZENG,WANG}] on Nd(O,F)BiS$_2$ report rather small electron doping levels of roughly 7 \% per Bi site, which is smaller than the high electron doping level, $x \sim$ 0.5, expected from theoretical studies. A low electron doping level, intrinsic structural instabilities[\onlinecite{LEE1}], and possible bismuth deficiencies[\onlinecite{YE}] in BiS$_2$ compounds also complicate our ability to compare results from different studies and combinations of those factors possibly yield such disparities in crystallographic and superconducting properties between studies. In order to advance our understanding of superconductivity in BiS$_2$-based compounds, these materials issues must be addressed.

\section{Concluding Remarks}
We have studied the effect of partial chemical substitution of yttrium for lanthanum in the superconducting LaO$_{0.5}$F$_{0.5}$BiS$_{2}$ system. We synthesized polycrystalline samples of La$_{1-x}$Y$_{x}$O$_{0.5}$F$_{0.5}$BiS$_{2}$ up to $x =$ 0.40 and observed a solubility limit near $x =$ 0.20. All samples crystallized in the CeOBiS$_2$-type structure. The physical properties of the system were investigated via electrical resistivity, dc and ac magnetic susceptibility, and specific heat measurements. We found a correlation between the lattice constant $c$, the La-O-La bond angle, and the critical temperature $T_c$. The chemical pressure induced by yttrium substitution for lanthanum is insufficient to induce the high-$T_c$ and/or the structural phase transitions observed in measurements of LaO$_{0.5}$F$_{0.5}$BiS$_2$ under applied pressure.

\begin{acknowledgments}
This work was supported by the US Air Force Office of Scientific Research - Multidisciplinary University Research Initiative under Grant No. FA 9550-09-1-0603 (superconductivity search), the US Department of Energy under Grant No. FG0204-ER46105 (characterization and physical properties measurements), and the National Science Foundation under Grant No. DMR 1206553 (low-temperature measurements).
\end{acknowledgments}

\bibliography{LaYOFBiS2}

\end{document}